\documentclass[letterpaper,twocolumn,10pt]{article}
\usepackage{main}

\usepackage{tikz}
\usepackage{amsmath}
\usepackage{enumitem}
\usepackage{filecontents}

\usepackage{booktabs}
\usepackage{multirow}
\usepackage{makecell}
\usepackage{array}

\usepackage{times}
\usepackage{fullpage}
\usepackage[english]{babel}
\usepackage{graphicx}
\usepackage{amsmath}
\usepackage{amssymb}
\usepackage{booktabs}
\usepackage{algorithm}
\usepackage{algorithmic}
\usepackage{url}
\usepackage{hyperref}
\usepackage{caption}
\usepackage{subcaption}
\usepackage{multirow}
\usepackage{xcolor}
\usepackage{graphicx}
\usepackage{placeins}
\usepackage{tabularx}
\usepackage{xurl}
\usepackage{subcaption}
\usepackage{marvosym}
\newcolumntype{C}{>{\centering\arraybackslash}X}

\newcommand{\sysname}{LatencyPrism} 

\begin{document}

\title{\Large \bf \sysname{}: Online Non-intrusive Latency Sculpting \\for SLO-Guaranteed LLM Inference}

\author{
  {\rm Yin Du}\footnotemark[1] \footnotemark[2] \\
  Alibaba Cloud Computing
  \and
  {\rm Jiayi Ren}\footnotemark[1] \\
  Alibaba Cloud Computing
  \and
  {\rm Xiayu Sun}\footnotemark[1] \\
  Alibaba Cloud Computing \\ Xi'an Jiaotong University
  \and
  {\rm Tianyao Zhou}\\
  Alibaba Cloud Computing
  \and
  {\rm Haizhu Zhou}\\
  Alibaba Cloud Computing
  \and
  {\rm Ruiyan Ma}\\
  Alibaba Cloud Computing
  \and
  {\rm Danyang Zhang}\\
  Alibaba Cloud Computing
}

\maketitle

\footnotetext[1]{These authors contributed equally to this work.}
\footnotetext[2]{Corresponding author. Email: \texttt{duyin.dy@alibaba-inc.com}}

\begin{abstract}
LLM inference latency critically determines user experience and operational costs, directly impacting throughput under SLO constraints. Even brief latency spikes degrade service quality despite acceptable average performance. However, distributed inference environments featuring diverse software frameworks and XPU architectures combined with dynamic workloads make latency analysis challenging. Constrained by intrusive designs that necessitate service restarts or even suspension, and by hardware-bound implementations that fail to adapt to heterogeneous inference environments, existing AI profiling methods are often inadequate for real-time production analysis.

We present \sysname{}, the first zero-intrusion multi-platform latency sculpting system. It aims to break down the inference latency across pipeline, proactively alert on inference latency anomalies, and guarantee adherence to SLOs, all without requiring code modifications or service restarts. \sysname{} has been deployed across thousands of XPUs for over six months. It enables low-overhead real-time monitoring at batch level with alerts triggered in milliseconds. This approach distinguishes between workload-driven latency variations and anomalies indicating underlying issues with an F1-score of 0.98. We also conduct extensive experiments and investigations into root cause analysis to demonstrate \sysname{}'s capability. Furthermore, we introduce the first LLM anomaly simulation toolkit to facilitate future research in robust and predictable inference systems.
\end{abstract}

\section{Introduction}

Large Language Models (LLMs) are now ubiquitous in production environments, supporting critical real-time applications such as conversational agents and code assistants~\cite{straits2025, hai2025}. As direct user-facing services, LLM inference requires not only traditional system uptime but also real-time responsiveness and stability~\cite{clone}, which are directly tied to the user's interactive experience. Recognizing this requirement, mainstream LLM providers have explicitly incorporated latency-related metrics into their SLAs. For example, OpenAI calculates p50 request latency over 5-minute intervals~\cite{openai_scale_tier}.

\label{para:lifecycle}
Reliability in this context extends beyond availability to \textit{latency stability}. However, distributed LLM serving suffers from significant latency jitter. This manifests as `Generation Stalls' where inter-token generation pauses for seconds~\cite{2403.02310}. Unlike traditional throughput-oriented tasks, such variance in Time-Between-Tokens (TBT) violates SLAs and degrades the user's perceived latency~\cite{llumnix, anyscale, 2507.09019, 2507.1015025}. A stuttering chatbot rapidly erodes user trust and disrupts the natural flow of interaction.

To guarantee SLAs, production operations typically adhere to a pipeline of \textit{Anomaly Detection $\rightarrow$ Root Cause Localization $\rightarrow$ Remediation}. However, with model sizes growing and the complexity of distributed clusters increasing, achieving timely and accurate anomaly detection presents a formidable challenge. Existing monitor tools, relying on aggregated metrics and static thresholds, lack the granularity and workload-awareness required for LLM inference. These limitations lead to alerting failures. Critically, the observability gap between low-level hardware behavior and high-level business logic fractures the link between detection and localization, forcing operators to resort to expensive resource over-provisioning or laborious manual debugging when ensuring performance stability~\cite{2305.15778}.

To mitigate these challenges, real-time latency characterization of production LLM inference is critical. Inference frameworks typically report end-to-end latency aggregated at the request level, thereby obscuring internal performance bottlenecks. Meanwhile, it conflates queue scheduling time with actual execution time, masking the true performance bottlenecks by confusing extrinsic scheduling overheads with intrinsic processing latency. We advocate for decoupling these components to focus on on-device model execution time per inference step, providing precise and actionable insights for diagnosis, resource scheduling, and optimization.

We propose \textbf{\sysname{}}, the first online, multi-platform, and full-stack performance tracking system tailored for production LLM inference. Its core design methodology focuses on continuous system health monitoring with minimal overhead, automatic anomaly alerting, and synchronized on-demand deep tracing. Designed for non-intrusive deployment, \sysname{} operates without code modifications or service restarts. Deployed on a cluster of thousands of XPUs for over six months, it has demonstrated significant operational value.

This paper makes the following contributions:
\begin{enumerate}
    \item \textbf{Non-intrusive Cross-Stack Semantic Profiling.} We introduce a unified computational view spanning models, frameworks, the OS, and heterogeneous hardware. By bridging high-level semantics with low-level hardware execution without service interruption, \sysname{} establishes a consistent basis for latency characterization.
    \item \textbf{Event-Driven Unified Temporal Modeling.} We leverage distributed coordination to achieve dynamic collaboration of collection policies and nanosecond-level time alignment across hundreds of nodes. Our elastic collection architecture activates data channels on demand. This ensures high throughput and zero data loss while providing precise and consistent temporal semantics.
    \item \textbf{Efficient Online Monitoring and Context Reconstruction.} \sysname{} employs a two-stage strategy: an always-on monitor ($<$0.5\% CPU overhead) that detects baseline deviations in real-time, and an on-demand deep tracer ($\sim$7\% overhead). This approach enables operators to reconstruct the full execution context of anomalies, striking a balance between visibility and overhead.
    \item \textbf{Comprehensive Validation.} We evaluate \sysname{} against several categories of real-world scenarios, including computation contention, bandwidth saturation, and throttling. To support this, we open-source the first comprehensive anomaly simulation suite specifically tailored for Large Model Distributed Inference. Experiments and deployment data show that \sysname{} achieves an F1-score of 0.98 in anomaly detection. The high-fidelity context data effectively supports downstream root cause analysis and significantly accelerates manual troubleshooting. 
\end{enumerate}

\section{Motivation}

As LLMs graduate from experimental environments to production deployment, the operational landscape exhibits several distinct characteristics that differentiate it from traditional training or offline inference scenarios (Table~\ref{tab:workload_comparison}). These include cost-driven multi-tenancy~\cite{wallaroo, 2508.20274}, complex software stacks utilizing advanced optimization techniques~\cite{runai}, and the increasingly diverse heterogeneous hardware deployments~\cite{bentoml}. These factors intertwine to create a highly dynamic and opaque system environment. In this context, inference performance degradation rarely manifests as simple, reproducible service crashes. Instead, it appears as sporadic, non-reproducible generation stalls.

As mentioned in §~\ref{para:lifecycle}, the unique performance jitter characteristics of LLM inference severely disrupts the first two stages of standard reliability life cycle, posing strict challenges to SLA assurance.

\begin{table}[!b]
\centering
\small
\caption{Comparison of workload characteristics between Production LLM Inference and traditional Training/Offline Inference. Production environments face unpredictable request lengths, fluctuating rates (tidal effects), complex hardware environments (colocation), and rapid software iteration.}
\resizebox{\columnwidth}{!}{
\begin{tabular}{lp{3cm}p{3.5cm}}
\toprule
\textbf{Dimension} & \textbf{Training/Offline} & \textbf{Production Inference} \\
\midrule
Request Size & Fixed Batch & Variable, unpredictable \\
Request Rate & Fixed & Tidal, unpredictable \\
Hardware & Exclusive & Complex (Colocation) \\
Model Switching & None & Frequent (Multi-model) \\
Software Env & Single & Complex, rapid updates \\
\bottomrule
\end{tabular}
}
\label{tab:workload_comparison}
\end{table}

\subsection{Blind Spots in Anomaly Detection under Dynamic Loads}

The first step is the accurate and timely detection of performance regression. However, the distinct workload characteristics of LLM inference cause traditional monitoring systems to fail, trapping operators in a dilemma between false negatives and false positives.

\textbf{Aggregation Metrics Mask Micro-Stalls (Risk of False Negatives).} Existing monitoring metrics typically focus on request-level end-to-end latency or system throughput. However, LLM inference is an iterative generation process based on tokens, where a single request involves hundreds of iterations. Sporadic generation stalls occurring during intermediate steps are easily smoothed out in aggregated metrics like average latency. Consequently, while dashboard monitors may indicate normal operation, the user's real-time interaction experience is already severely compromised. Operators fail to perceive the anomaly immediately, missing the signal at its source.

\textbf{Load Fluctuations Invalidate Static Thresholds (Risk of False Positives).} Even when latency fluctuations are monitored, determining whether they constitute an anomaly is difficult. Unlike traditional microservices, the computation time for LLM inference is highly correlated with input/output token length and cache hit rates. In production environments, the high variance in request length results in a wide distribution of normal processing times. Traditional alert strategies based on static thresholds fail in this scenario. A low threshold triggers excessive false alarms for normal long-text requests, causing alert fatigue; conversely, a high threshold misses performance regressions in short-text requests. Existing systems lack workload awareness, making it difficult for operators to distinguish between \textbf{reasonable increases} due to higher workload and \textbf{performance degradation} caused by system anomalies.''

\textbf{Context Loss Due to Monitoring Lag.} A more critical issue is timeliness. Traditional monitoring systems often rely on polled metrics or log aggregation, which suffer from minute-level lags~\cite{2305.15778}. However, the lifecycle of an inference request is extremely short, and generation stalls are often caused by transient resource contention, such as PCIe bandwidth saturation or instantaneous throttling. By the time lagging monitoring metrics trigger an alert, the anomalous request has concluded, and the ephemeral context has vanished. This post-hoc non-reproducibility dictates that offline analysis is insufficient, necessitating an \textit{Online Monitoring and Reconstruction} mechanism to capture snapshots within the millisecond-level window of the anomaly.

\subsection{Difficulty in Latency Attribution}

Even if the monitoring system successfully captures an anomaly signal, existing monitoring data is often uninterpretable. The disconnection between software and hardware semantics prevents the decomposition of end-to-end latency into causal components, making it difficult to pinpoint the source of the delay. This is primarily due to two technical gaps.

\textbf{Missing Cross-Stack Semantics.} LLM inference is a process of asynchronous coordination between CPU scheduling (Python/C++) and GPU computation. High-level application logs only record the flow of business logic, while low-level hardware monitoring only records the utilization of physical resources. Lacking a unified context identifier to bridge these layers, operators cannot correlate latency spikes with resource states. Consequently, for example, it becomes indistinguishable whether a stall originates from Python runtime overhead, GPU kernel blocking, or PCIe bandwidth saturation.

\textbf{Temporal Misalignment.} Software logs and hardware metrics operate on divergent timebases and granularities (e.g., event-driven vs. periodic sampling). This mismatch creates severe observational blind spots. For example, a request-level latency spike seen by the application often fails to align with any specific hardware transient in the monitoring window. The logical slowness perceived by software and the transient instability of hardware simply do not align on the timeline. As a result, even laborious manual correlation fails to reconstruct the root cause context.

This attribution gap exacts a heavy toll on operational efficiency. Operators are forced to choose either spend significant effort manually parsing massive logs~\cite{2305.15778} or resort to conservative resource strategies, relying on expensive resource over-provisioning or service restarts to alleviate symptoms. Both methods result in low utilization of costly computing resources~\cite{2509.18101}.

\subsection{The Trade-off Dilemma in Existing Observability}
\label{sec:tradeoff}
These challenges dictate the design requirements for an effective collection system. However, the existing tools fails to simultaneously satisfy three non-negotiable constraints for production inference: Non-intrusiveness, Low Overhead, and Full-Stack Semantics. Operators are consequently trapped in an observability trilemma:

\begin{itemize}
    \item \textbf{Depth vs. Availability.} Obtaining precise underlying GPU states (e.g., kernel-level duration) typically requires intrusive profilers provided by hardware vendors. These tools often demand service restarts for process takeover and are accompanied by significant prohibitive overhead, making them unsuitable for online production use.
    \item \textbf{Semantics vs. Visibility.} While framework-integrated tools can understand business logic at the Python level, their visibility is confined within the framework, unable to perceive external interference in multi-tenant environments.
    \item \textbf{Transparency versus Blind Spots.} Emerging eBPF technology guarantees low overhead and non-intrusive attachment but currently suffers from inherent visibility deficits regarding GPU internals, making it difficult to support deep performance diagnosis.
\end{itemize}

\subsection{Summary of the Observability Gap}

As summarized in Table~\ref{tab:tool_comparison}, current tooling paradigms fail to simultaneously address the workload characteristics, the attribution complexity, and the operational constraints of LLM inference.

\begin{table*}[t]
\centering
\small
\caption{Comparison of \sysname{} with existing profiling and reliability paradigms.}
\begin{tabularx}{\textwidth}{@{} l c C C C C @{}} 
\toprule
\textbf{Category} & \textbf{Target} & \textbf{Online} & \textbf{Non-} & \textbf{Latency} & \textbf{Cross-Stack} \\
\textbf{(Representative Tools)} & \textbf{Workload} & \textbf{Safe} & \textbf{Intrusive} & \textbf{Sculpting} & \textbf{Context} \\
\midrule
\textbf{Vendor Profilers} & General & No & No & Manual & Hardware \\
(Nsight) & & & (Process Launch) & & Only \\
\midrule
\textbf{Framework Tools} & General & No & No & Manual & Software\\
(Torch Profiler) & & & (Modify Code) & & Only \\
\midrule
\textbf{eBPF Observability} & General & \textbf{Yes} & \textbf{Yes} & N/A & System \\
(DeepFlow) & (CPU-bound) & & & & Only \\
\midrule
\textbf{Training Reliability} & \textbf{Training} & N/A & No & N/A & Software \\
(Mycroft, XPUTimer) & \textbf{Only} & & (Modify Libs) & & Only \\
\midrule
\textbf{\sysname{}} & \textbf{General} & \textbf{Yes} & \textbf{Yes} & \textbf{Auto} & \textbf{Unified} \\
 (Ours) & (Inference Motivated) & & & & \textbf{(HW + SW)} \\
\bottomrule
\end{tabularx}
\label{tab:tool_comparison}
\end{table*}

\subsection{Challenges}
Building a production-grade profiling system for LLM inference entails overcoming four fundamental challenges:
\begin{itemize}
    \item \textbf{Constraints on Intrusiveness and Overhead.} This is the prerequisite for any production deployment. In production inference, adhering to strict SLAs precludes service restarts or downtime for instrumentation purposes. Thus, the system must be non-intrusive and restart-free.  Furthermore, latency-sensitive inference is highly susceptible to the observer effect: indiscriminate event collection competes for resources, introducing contention that might mask true bottlenecks~\cite{overheadanalysis}. This necessitates an adaptive collection strategy that that maximizes information gain while minimizing overhead.
    \item \textbf{Effective Full-Stack Observability.} Effective diagnosis requires correlating high-level Python business logic with low-level hardware metrics (e.g., PCIe throughput, GPU frequency). The core difficulty lies in temporal and semantic alignment: mapping asynchronous GPU kernel executions back to CPU-side application contexts across disparate granularities (coarse metrics vs. fine-grained traces). Moreover, the diversity of hardware backends requires a unified abstraction layer to mask heterogeneous complexity.
    \item \textbf{In-Situ Context Capture for Transient Anomalies.} Inference workloads are characterized by short lifecycles and high concurrency. Traditional anomaly detection, relying on static thresholds derived from long-running training tasks, fails to generalize, leading to high false positives. Crucially, inference anomalies are often transient and ephemeral. Without in-situ snapshotting capabilities triggered precisely at the onset of an anomaly, the execution context vanishes instantly, rendering post-hoc reconstruction impossible.
    \item \textbf{Lack of Global Coordination.} In distributed inference, performance degradation is often the result of coupled inter-node network congestion and intra-node resource contention. Existing tools provide only siloed, single-node views, lacking the ability to trigger synchronized tracing across the cluster. This fragmentation forces operators to manually correlate TB-scale logs without a globally consistent timeline, turning fault localization into an inefficient guessing game~\cite{rootcause, rootcause2}.
\end{itemize}

\section{Design}

\begin{figure}[!b]
    \centering
    \includegraphics[width=\columnwidth]{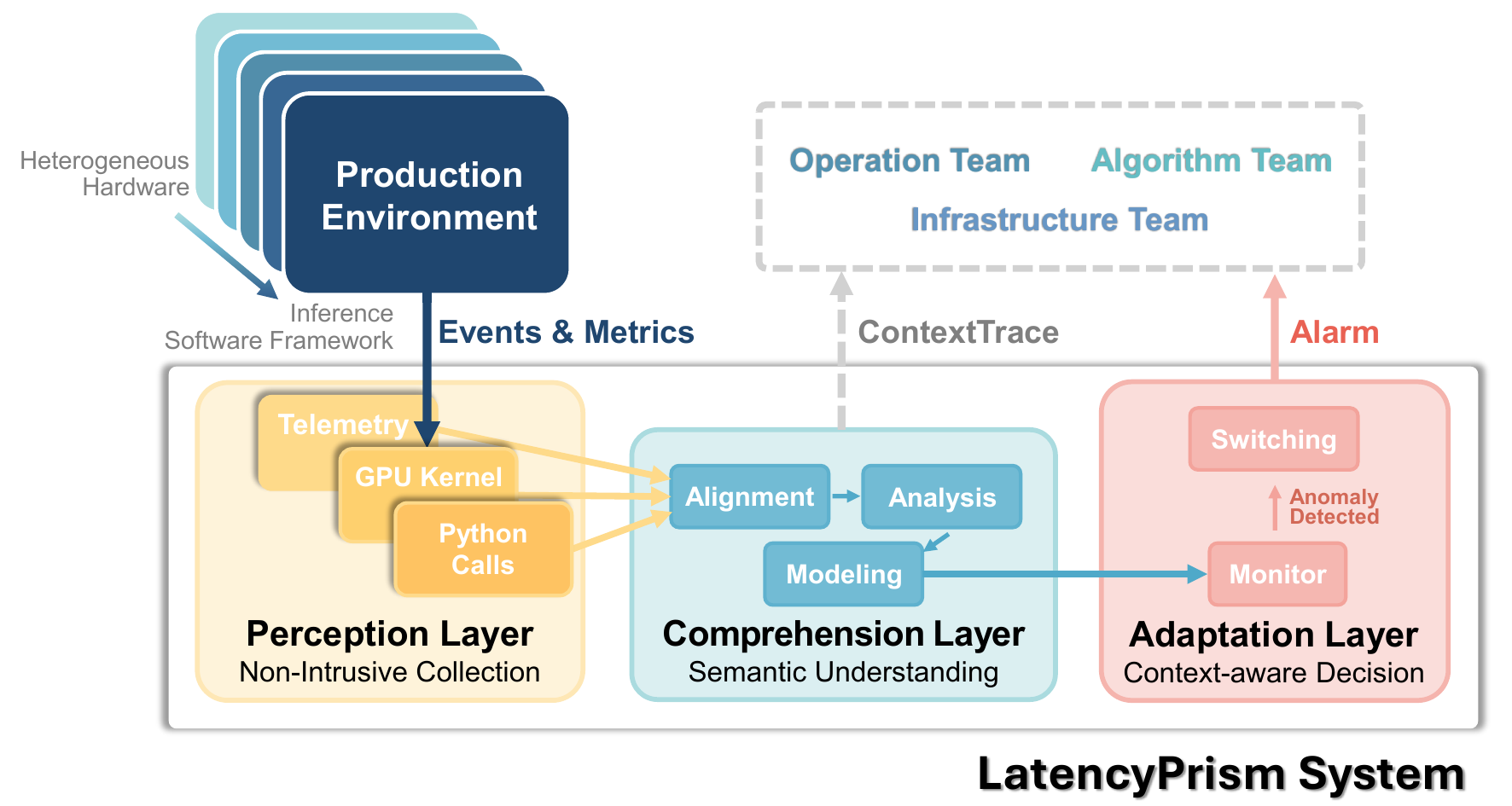}
    \caption{\textbf{Architecture Overview of \sysname{}.} The system consists of three layers: (1) The Perception Layer collects non-intrusive runtime events and telemetry; (2) The Comprehension Layer aligns cross-stack events and constructs workload-aware baselines; (3) The Adaptation Layer monitors latency residuals, triggering alerts providing deep-dive contexts when anomalies detected.}
    \label{fig:architecture}
\end{figure}

We present \sysname{}, an online, dynamic system tailored for LLM inference. Figure~\ref{fig:architecture} illustrates the system architecture. To navigate the trade-offs discussed in §~\ref{sec:tradeoff}, \sysname{} adheres to four guiding principles: Lightweight Operation and Zero-Instrumentation (addressing transparency constraints), supported by Full-Stack Consistency and Broad Compatibility (addressing observability gaps). The system is actualized through three synergistic layers: Perception, Comprehension, and Adaptation.

\subsection{Design Principles}
\textbf{Lightweight.} The system supports configurable collection granularity. Users can freely configure collection plugins according to their needs, toggling collection at various levels, down to specific event types or metrics. Furthermore, the system supports dynamic plugging and configuration of collection tools. Trace data is exported via shared memory to an independent processing process to avoid buffer accumulation within the target container. Upon completion of collection, dynamically injected probes are automatically unloaded to minimize interference.

\textbf{Zero Intrusion.} We acknowledge that completely zero-intrusion observation is an idealized goal. Observing the complete state of a process inevitably requires some interaction with its runtime behavior. In this paper, Zero Intrusion refers to transparency at the \textit{user operation level}. All probes attach automatically during runtime. Whether initiating a new collection or changing a configuration, users are not required to modify code, adjust launch commands, or restart services. While we utilize mechanisms like \texttt{ptrace} and CUPTI, the user's application logic remains untouched.

\textbf{Full-Stack Consistency.} Collection must be comprehensive, covering the entire software stack from hardware, the operating system, and the runtime to the AI framework layer. The system synchronizes the collection of micro-events (e.g. python function calls, CUDA kernels) and macro-metrics (utilization, bandwidth, temperature, frequency) to construct an end-to-end performance profile.

\textbf{Compatibility.} To accommodate the complex technology stack of inference scenarios, we prioritize compatibility across hardware platforms and software stacks. In addition to NVIDIA GPUs, We abstract the XPU plugin layer, allowing rapid compatibility implementation for other hardware with similar driver-level capabilities. Regarding the software stack, we provide extensive support for Python-based frameworks. For other languages, while detailed information comparable to Python may be unavailable, kernel event streams remain accessible for analysis. Although motivated by LLM inference, \sysname{} supports various models including LLM training, VLMs, and Diffusion models.

\subsection{System Components}

\sysname{} serves not only as a data collection system but also as a semantic understanding engine for LLM inference. Its core consists of three collaborative components: the Perception Layer, the Comprehension Layer, and the Adaptation Layer.

\subsubsection{Perception Layer: Non-Intrusive Cross-Platform Collection}
This layer forms the sensory foundation, addressing the challenges of data acquisition and transparency. It constructs four collaborative observation planes to achieve full-link observability without code modification or service restarts:
\begin{itemize}
    \item \textbf{CPU and System Behavior Tracking.} Leveraging kernel-level observability technology to monitor operating system scheduling, system calls, and network I/O at a fine grain, reconstructing the complete CPU execution path.
    \item \textbf{Transparent Runtime Instrumentation.} Automatically collecting Python function call stacks, model metadata (e.g., model structure, parameter size), and key operation markers (e.g., NVTX), without user intervention.
    \item \textbf{GPU Fine-Grained Monitoring.} Penetrating the GPU driver layer to capture CUDA kernel execution, memory copies, and stream scheduling, providing nanosecond-level timelines aligned with CPU events.
    \item \textbf{System Telemetry.} Synchronously collecting macro resource states such as CPU/GPU utilization, memory bandwidth, throughput, device temperature, and operating frequency, fusing micro-behavior with system-level performance profiles.
\end{itemize}
To support large-scale clusters, this layer enables one-click triggering of cross-node distributed collection tasks, automatically aggregating multi-node trace data. Additionally, an asynchronous task controller and multi-point sampling strategy ensure alignment of events from different levels within a unified coordinate system.

\subsubsection{Comprehension Layer: Automatic Parsing of Inference Semantics}
Data reported by the collection layer is typically discrete, stateless time series (e.g., instantaneous GPU utilization or independent kernel events). This layer transforms low-level data into business-valuable information through the following mechanisms:

\textbf{Cross-Stack Semantic Alignment.} The system employs timestamp calibration and causal correlation algorithms to map application-layer Python function calls (business intent) to underlying GPU kernel execution (hardware implementation) accurately.

\textbf{Cyclic Iteration Recognition.} Targeting the autoregressive generation characteristics inherent in LLM/VLM inference, this layer automatically identifies key functions within the inference process. It segments the continuous event stream into independent, semantically clear Batch granularities. This allows operators to analyze specific time windows rather than endless timelines.

\begin{figure}[!b]
\centering
\includegraphics[width=\columnwidth]{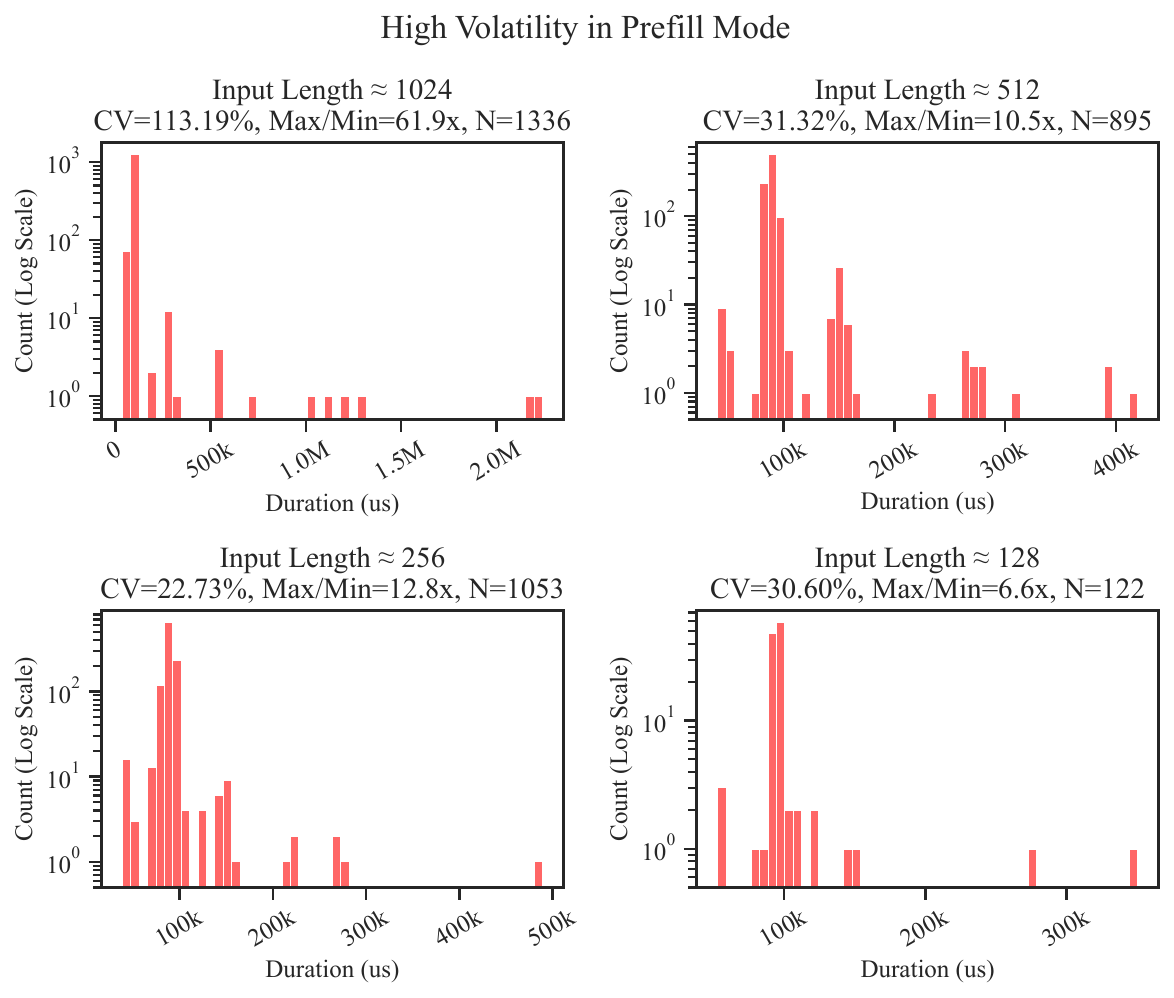}
\caption{Distribution of Prefill stage duration in a production environment. Even with fixed input lengths, execution time fluctuates drastically with a long tail, primarily due to KV Cache hit rate variations.}
\label{fig:prefill_hist}
\end{figure}

\textbf{Stage Awareness (Prefill vs. Decode).} The system intelligently classifies the inference phase of each batch. \sysname{} prioritizes the Decode stage for anomaly detection. Prefill latency is inherently volatile: advanced optimizations like PagedAttention~\cite{vllm} and RadixAttention~\cite{sglang} cause execution times to fluctuate wildly depending on KV-cache layout and hit rates, even for identical input lengths (Figure~\ref{fig:prefill_hist}). Conversely, these optimizations are critical for production systems and cannot be easily disabled. This variance makes physical baseline modeling for Prefill prone to noise, whereas Decode latency is more stable. Meanwhile, from a user-experience perspective, the Decode phase latency governs the perceived fluidity of the output stream, while variations in the initial Time-to-First-Token (TTFT) are generally better tolerated~\cite{tokenflow}.

\begin{figure*}[!t]
    \centering
    \begin{subfigure}[b]{0.39\linewidth}
        \centering
        \includegraphics[width=\linewidth]{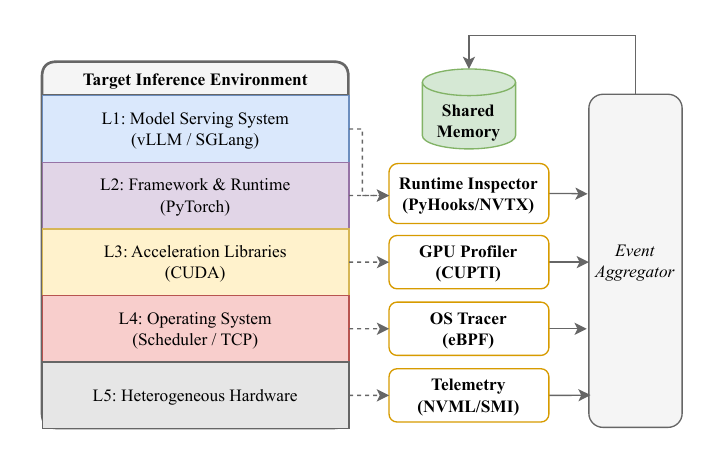}
        \caption{\textbf{Perception Layer}}
        \label{fig:perception}
    \end{subfigure}
    \hfill
    \begin{subfigure}[b]{0.59\linewidth}
        \centering
        \includegraphics[width=\linewidth]{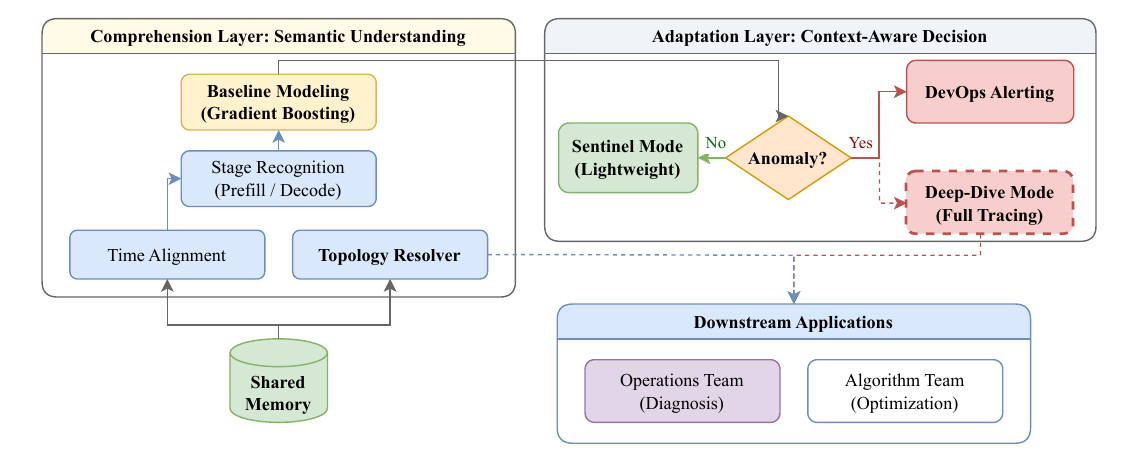}
        \caption{\textbf{Comprehension \& Adaptation Layer}}
        \label{fig:decision}
    \end{subfigure}
    
    \caption{\textbf{Detailed implementation workflow of \sysname{}.} 
    (a) Maps the heterogeneous inference environment (L1-L5) to non-intrusive collectors via Shared Memory. 
    (b) Illustrates the pipeline from trace alignment to GBDT-based anomaly detection. The system flows from left (collection) to right (decision).}
    \label{fig:implementation_flow}
\end{figure*}

\textbf{Context-Aware Baseline Modeling.} The system recognizes that latency is not an absolute value but a function of workload. This layer parses the workload context of the current Batch in real-time, combining historical data to build dynamic physical expectation baselines. This enables the system to distinguish between reasonable load increases and abnormal performance regression.

\subsubsection{Adaptation Layer: Context-Based Anomaly Detection and Response}
This layer balances collection overhead with diagnostic depth. To achieve both low overhead and high-fidelity diagnosis in production, \sysname{} adopts a dual-mode monitoring architecture. While not explicitly dependent on user-defined SLO thresholds, its anomaly detection mechanism naturally serves SLO assurance goals by identifying performance degradation events likely to violate service levels.

\textbf{Sentinel Mode.} The system defaults to this low-overhead mode, collecting only workload metadata (e.g., Input Tokens, Batch Size) and a minimal set of key application-layer events. This mode utilizes the relatively deterministic nature of LLM inference computation to build a theoretical expected duration baseline based on real-time load. Its core objective is to decouple ``load variation'' from ``performance regression''—distinguishing in real-time whether a latency increase stems from a heavier task (normal) or a system anomaly (failure).

\textbf{Deep-Dive Mode.} When Sentinel Mode detects a significant deviation of actual duration from the theoretical baseline, the system automatically flags an anomaly. Based on user-configured policies, it issues alerts and dynamically triggers comprehensive collection (e.g., enabling Python Probes or GPU kernel Traces). This mechanism ensures negligible interference during system health while precisely capturing context during anomalies for subsequent analysis.

\section{Implementation}

\subsection{Dynamic Non-Intrusive Online Tracing}

\subsubsection{Collection Methodology}
\sysname{} implements a multi-level tracing system. At the application level, we employ a lightweight probe inspired by PyTorch Dynamo that leverages ptrace to dynamically instrument \texttt{PyFrameObject}. This approach utilizes ptrace to non-intrusively hooks the creation and destruction processes of Python virtual machine frames at runtime, capturing high-fidelity execution context without requiring code modification or process restarts. It extracts metadata such as function name, records call timing with nanosecond precision, and structurally parses arguments and return values. Moreover, it automatically identifies model metadata in mainstream AI frameworks (e.g., vLLM, SGLang) and inserts semantically labeled NVTX or ROCTX range events during key function calls, achieving semantic alignment between GPU kernels and Python logic. The probe also manages reference counting meticulously to avoid interfering with the target process's garbage collection.

Beyond the application layer, the system integrates mature solutions at other levels: eBPF is used to trace CPU scheduling, network I/O, and system calls in kernel space; CUPTI or ROCm GPUPerfAPI collects underlying GPU activities; for other heterogeneous devices, corresponding \texttt{-smi} tool outputs are parsed. The system further aggregates micro-events with macro-metrics to achieve full-stack observability.

\subsubsection{Multi-Dimensional Data Correlation}
\begin{figure*}[t]
\centering
\includegraphics[width=\textwidth]{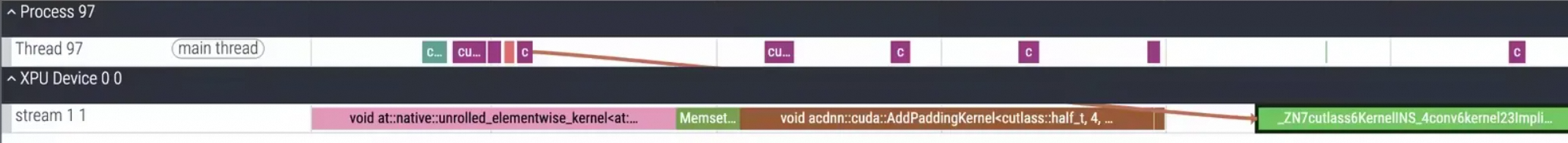}
\caption{Trace view demonstrating cross-stack semantic alignment. \sysname{} aligns CUDA Runtime API calls with underlying GPU kernel executions (e.g., \texttt{AddPaddingkernel}) on a unified timeline.}
\label{fig:trace_view}
\end{figure*}

Although GPU event timestamps, CPU-side eBPF probe timestamps, and Python probe timestamps originate from different sources, the \sysname{} backend data processing pipeline utilizes multiple synchronization beacons and time calibration registrations embedded during collection to align these heterogeneous data sources precisely on a unified timeline. Furthermore, every activity on the GPU (e.g., kernel execution or memory copy) is linked to the CUDA Runtime API or Driver API call that triggered it on the CPU side. This visualizes the entire process from API invocation to driver submission and final GPU execution, as shown in Figure~\ref{fig:trace_view}.

\subsubsection{Distributed Topology Resolution}
\label{sec:topology}
In large-scale distributed LLM inference, collaborative operations such as AllReduce communication in tensor parallelism or cross-GPU synchronization of KV Cache are typically abstracted in application code via logical identifiers (e.g., NCCL's \texttt{commHash}). Observable low-level events (e.g., GPU kernel, NVLink traffic, network packets), however, bind to physical resources (e.g., \texttt{node-07/gpu2}). If this mapping is missing, even full-stack traces cannot answer critical questions like ``Between which two physical GPUs is this slow AllReduce operation occurring?''

To bridge this gap, \sysname{} parses \texttt{commHash}, \texttt{rank}, \texttt{node}, and \texttt{hostname} parameters in NCCL events to dynamically construct a global mapping table of $(\texttt{commHash}, \texttt{rank}) \rightarrow (\texttt{node}, \texttt{device})$, thereby precisely associating logical communication relationships with the physical topology.

\subsection{Online Monitoring}

\subsubsection{Iteration Cycle Recognition}
\sysname{} automatically delineates inference cycles based on Python call stacks. The system scans high-level Python function calls in the trace data. By analyzing call frequency and duration stability, it automatically identifies anchor functions that best represent a complete inference iteration cycle. Using these anchors as boundaries, the continuous event stream is segmented into independent cycles. In scenarios requiring extremely low overhead or non-Python drivers, we support cycle recognition with lower precision via frequency domain analysis of GPU kernel or User Probe event streams.

To distinguish between Prefill and Decode stages, the system analyzes multiple indicators. It matches predefined key sub-event lists, such as \texttt{forward\_prefill} and \texttt{process\_batch\_result\_decode}, and utilizes hooked parameter information like \texttt{forward\_mode}. Furthermore, it leverages temporal characteristics. Prefill stages are identified by their potentially longer preceding idle times and significantly greater duration compared to the brief, closely following Decode cycles. This approach enables cycle recognition at token granularity for LLMs and iteration granularity for VLMs.

\subsubsection{Predictive Modeling and Feature Selection}
To achieve high-precision anomaly detection, we construct a baseline model with workload as input and latency as output: $Latency = f(Workload)$. Taking SGLang framework as an example, an iteration consists of three parts, \texttt{get\_next\_batch\_to\_run}, \texttt{run\_batch} and \texttt{process\_batch\_result}. Scheduling-related overhead (\texttt{get\_next\_batch\_to\_run}) is intentionally excluded from the modeling scope for three reasons: (1) Data Validity: monitoring scheduling during idle periods introduces noise; (2) Focus: performance anomalies rarely manifest solely as scheduling delays; and (3) Decoupling: scheduling latency depends on global system state rather than the current batch's workload.

The Decode stage theoretically exhibits linear complexity $O(B \cdot K) + O(B)$, where $B = \text{Batch Size}$ and $K = \text{KVCache} = \text{Input Length} + \text{Current Output Length}$.
Based on the cycle segmentation results, we can extract the runtime of a single batch. By collecting data during normal operation combined with load information, we can model the theoretical single-batch runtime. However, a critical challenge exists in actual deployments: in production environments, inference frameworks universally enable Overlap optimization, introducing non-linear logic of $\max(T_{GPU\_compute}, T_{comm\_and\_process})$. Experiments show that this inflection point behavior causes simple linear physical models to fail. Therefore, we employ Gradient Boosting Decision Trees (GBDT) as the baseline predictor. GBDT can fit the non-linear inflection points introduced by Overlap extremely well.

To capture memory bandwidth bottlenecks in the Decode stage, we construct physical features based on hardware principles. The core overhead of Decode operators lies in reading the Key-Value matrices of all historical tokens. We define $L_{real} = L_{in} + L_{out}$ and construct the feature $W_{kv} = B \times L_{real}$. This feature directly maps to GPU memory bandwidth pressure. We explicitly exclude certain statistical features during feature selection. Although introducing these features might improve performance data by capturing implicit relationships, they would undermine model interpretability.

\begin{figure*}[t]
    \centering
    \begin{subfigure}[b]{0.32\textwidth}
        \centering
        \includegraphics[width=\linewidth]{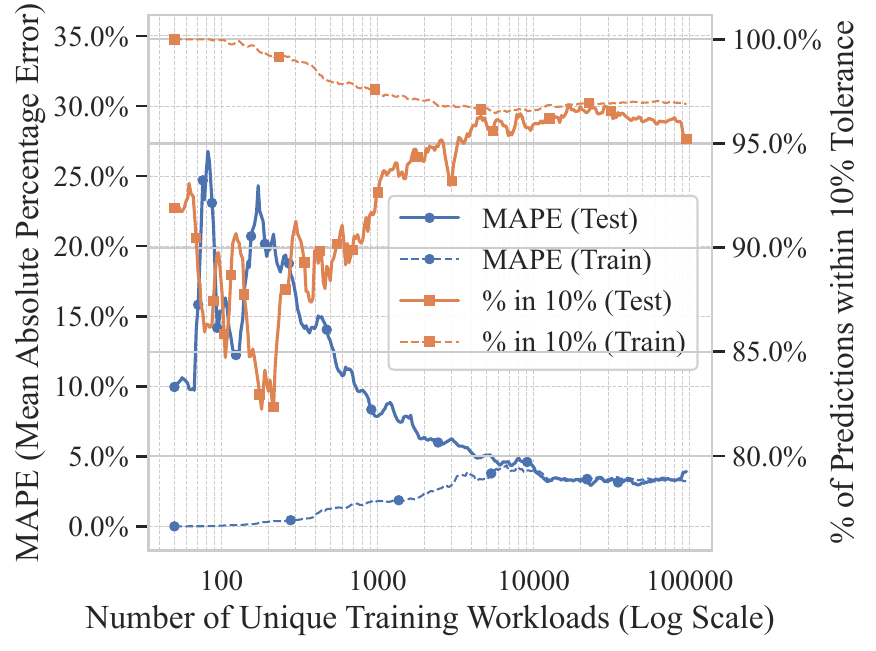}
        \caption{\textbf{Unseen Workloads}}
        \label{fig:curve_median}
    \end{subfigure}
    \hfill
    \begin{subfigure}[b]{0.32\textwidth}
        \centering
        \includegraphics[width=\linewidth]{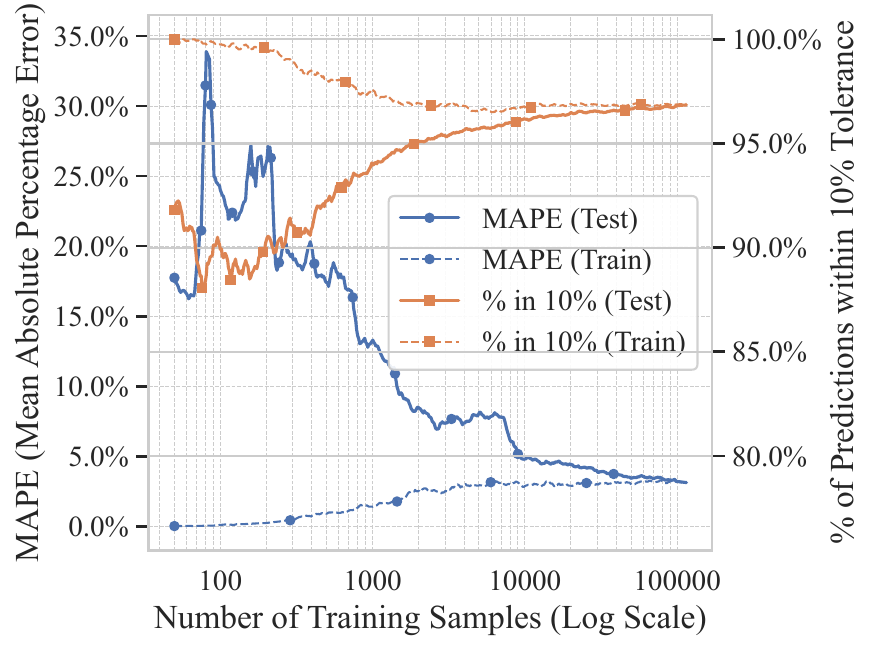}
        \caption{\textbf{Production Noise}}
        \label{fig:curve_raw}
    \end{subfigure}
    \hfill
    \begin{subfigure}[b]{0.32\textwidth}
        \centering
        \includegraphics[width=\linewidth]{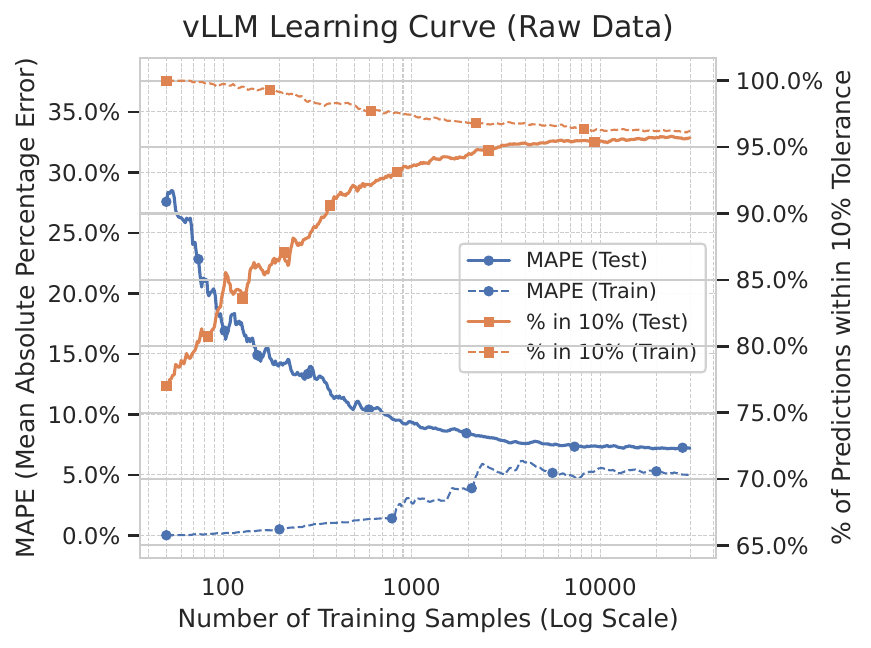}
        \caption{\textbf{Cross-Stack}}
        \label{fig:curve_vllm}
    \end{subfigure}
    
    \caption{\textbf{Model Convergence Analysis.} The model achieves rapid stabilization across three distinct scenarios: (a) unseen workloads in SGLang; (b) raw production noise without denoising; and (c) a cross-stack environment (vLLM, DeepSeek-70B), demonstrating robust generalization capabilities.}
    \label{fig:convergence_analysis}
\end{figure*}

\subsubsection{Alert Trigger and Comprehensive Collection}
We employ control chart theory to monitor prediction residuals. We define the Positive Prediction Error (PPE) as the monitoring metric $E_t = \max(0, \frac{Y_t - \hat{Y}_t}{Y_t + \epsilon})$, focusing only on cases where actual time significantly exceeds predicted time (i.e., performance degradation). To smooth transient noise, we introduce a sliding window (Window Size $W$) to calculate the moving average error $\bar{E}_t$. The system dynamically calculates the Upper Control Limit (UCL) based on the residual distribution of the training set:
\begin{equation}
UCL_{dynamic} = \min(\mu_{train} + k \cdot \sigma_{train}, \theta_{max})
\end{equation}
where $\mu$ and $\sigma$ are the mean and standard deviation of the training set errors, $k=3$ is the sigma coefficient, and $\theta_{max}$ is the empirically set maximum tolerance. When $\bar{E}_t > UCL_{dynamic}$, the system triggers an anomaly alert and initiates comprehensive collection according to user configuration.

To evaluate \sysname{} under diverse and realistic stress conditions, we developed an anomaly simulation suite. These simulations are organized according to the fundamental components of a modern, accelerator-equipped Von Neumann architecture, allowing us to systematically inject performance regressions by targeting specific hardware resources.

\begin{table*}[t]
\centering
\caption{Performance and Interpretability Analysis of Feature Engineering Strategies.}
\label{accuracy_comparison}
\resizebox{\textwidth}{!}{
\begin{tabular}{ll c c l l}
\toprule
\textbf{Strategy} & \textbf{Model} & \textbf{$R^2$} & \textbf{MAPE} & \textbf{Dominant Features (Top-weighted)} & \textbf{Analysis \& Interpretation} \\
\midrule
Full Feature & GBDT & 0.989 & 1.53\% & \texttt{post\_MaxInLen}, \texttt{post\_FwdMode} & \textbf{Pipeline Coupling.} Overfits to SGLang's specific overlap logic rather than execution physics. \\
Full Feature & Polynomial & 0.865 & 5.98\% & \texttt{BatchSize}$\times$\texttt{FwdMode} (Coeff $\approx -1.3 \times 10^6$) & \textbf{Multicollinearity.} Massive, opposing coefficients indicate mathematical instability. \\
\midrule
\textbf{Feature Eng.} & \textbf{GBDT (Ours)} & \textbf{0.963} & \textbf{6.06\%} & \textbf{\texttt{Workload\_KV} ($B \cdot L$), \texttt{Batch}} & \textbf{Physical Causality.} Correctly identifies memory bandwidth bottleneck ($O(B \cdot L)$). \\
Feature Eng. & Polynomial & 0.056 & 35.3\% & \texttt{BatchSize} & \textbf{Model Mismatch.} Fails to capture the non-linear inflection point of Overlap. \\
\bottomrule
\end{tabular}
}
\end{table*}

\begin{itemize}
    \item \textbf{Processing Units:} We simulate contention and performance degradation within the primary computational components of the system.
    \begin{itemize}
        \item \textbf{CPU (Central Processing Unit):} To stress the CPU's Arithmetic Logic Units (ALUs), we utilize AVX-512 saturation with dense floating-point loops. To simulate thermal throttling or power-saving modes, we enforce Dynamic Voltage and Frequency Scaling (DVFS) constraints using \texttt{cpupower}.
        \item \textbf{GPU (Graphics Processing Unit):} We generate contention on the GPU's Streaming Multiprocessors (SMs) by launching high-intensity General Matrix Multiply (GEMM) kernels. Additionally, we simulate power-limited or thermally constrained scenarios by locking the graphics and memory clocks to their minimum supported frequencies via \texttt{nvidia-smi -lgc}.
    \end{itemize}

    \item \textbf{Memory System:} We target the memory hierarchy to simulate Memory Thrashing and Fragmentation. We induce a state of high paging activity where the system excessively swaps data between RAM and disk, leading to severe performance degradation due to high I/O wait times. It can also break available memory into small, non-contiguous blocks after numerous allocation and deallocation cycles.

    \item \textbf{Interconnects and Data Bus:} We create bottlenecks on the data pathways connecting the system's components.
    \begin{itemize}
        \item \textbf{GPU-to-GPU Interconnect (NVLink):} We saturate the high-speed NVLink bus through continuous peer-to-peer memory transfers, simulating heavy multi-GPU workloads.
        \item \textbf{CPU-to-GPU Interconnect (PCIe):} We saturate the PCIe bus by programmatically disabling peer-to-peer (P2P) communication. This forces data transfers between GPUs to be relayed through host memory (Device $\rightarrow$ Host $\rightarrow$ Device), creating a significant bottleneck.
        \item \textbf{Bus Contention:} We simulate contention from co-located devices, such as multiple GPUs and high-speed network interface cards (NICs) competing for limited PCIe lane bandwidth. This scenario mimics resource competition in densely populated server nodes, where one device's high traffic can starve another.
    \end{itemize}
\end{itemize}

\begin{table*}[t]
\centering
\caption{Performance and Interpretability Analysis of Feature Engineering Strategies.}
\label{accuracy_comparison}
\resizebox{\textwidth}{!}{
\begin{tabular}{ll c c l l}
\toprule
\textbf{Strategy} & \textbf{Model} & \textbf{$R^2$} & \textbf{MAPE} & \textbf{Dominant Features (Top-weighted)} & \textbf{Analysis \& Interpretation} \\
\midrule
Full Feature & GBDT & 0.989 & 1.53\% & \texttt{post\_MaxInLen}, \texttt{post\_FwdMode} & \textbf{Pipeline Coupling.} Overfits to SGLang's specific overlap logic rather than execution physics. \\
Full Feature & Polynomial & 0.865 & 5.98\% & \texttt{BatchSize}$\times$\texttt{FwdMode} (Coeff $\approx -1.3 \times 10^6$) & \textbf{Multicollinearity.} Massive, opposing coefficients indicate mathematical instability. \\
\midrule
\textbf{Feature Eng.} & \textbf{GBDT (Ours)} & \textbf{0.963} & \textbf{6.06\%} & \textbf{\texttt{Workload\_KV} ($B \cdot L$), \texttt{Batch}} & \textbf{Physical Causality.} Correctly identifies memory bandwidth bottleneck ($O(B \cdot L)$). \\
Feature Eng. & Polynomial & 0.056 & 35.3\% & \texttt{BatchSize} & \textbf{Model Mismatch.} Fails to capture the non-linear inflection point of Overlap. \\
\bottomrule
\end{tabular}
}
\end{table*}

\section{Evaluation}

\subsection{Experimental Setup}
\textbf{Hardware and Models.} Unless otherwise stated, we evaluated \sysname{} on NVIDIA A100 GPUs using SGLang v0.5.4 and the Qwen3-32B model.

\textbf{Data Scope and Granularity.} Data was captured with all raw traces formatted for compatibility with standard visualization tools like Perfetto.
\begin{itemize}
    \item \textbf{Sentinel Mode Evaluation:} The workload distribution spanned a wide range to mimic production diversity: batch sizes of 1–512, input lengths of 1–2048, and output lengths of 1–512.
    \item \textbf{Deep Dive Mode Evaluation:} We recorded high-density traces for each anomaly scenario. Each trace encompasses multiple concurrent threads and logs millions of fine-grained events, providing the statistical basis for our exploratory root cause localization.
\end{itemize}

\subsection{Sentinel Mode Assessment}

\textbf{Overhead.} \sysname{} incurs negligible overhead, degrading throughput by less than 0.5\% and latency by less than 0.1\%, making it suitable for production deployment.

\textbf{Model Adaptability and Robustness.} We conducted three distinct evaluations to assess the model's performance:

\begin{itemize}
    \item \textbf{Generalization to Unseen Workloads.} To verify the model's ability to migrate between configurations, we split the dataset by unique workload types (i.e., the test set contained workloads never seen during training). As shown in Figure~\ref{fig:curve_median}, the results confirm that the model learns intrinsic performance characteristics, enabling rapid convergence on new configurations with minimal samples.

    \item \textbf{Robustness to Production Noise.} To simulate a real-world production environment, we used raw, noisy data with a completely random split, without segregating workloads. As shown in Figure~\ref{fig:curve_raw}, even without denoising, the model stabilizes within approximately 1,000 samples (equivalent to minutes of traffic), achieving a prediction error of $<10\%$ for over 90\% of samples. This demonstrates the model's robustness against production noise.

    \item \textbf{Cross-Stack Generalizability.} To confirm that our modeling strategy is not overfitting to specific framework implementations, we conducted an additional evaluation on vLLM v0.10.0 serving DeepSeek-R1-Distill-Llama-70B with Tensor Parallelism(TP) size 4. As shown in Figure~\ref{fig:curve_vllm}, \sysname{} successfully adapts to the distinct scheduling logic of vLLM and the computational patterns of the 70B model. The predictor converges rapidly, validating that our physically-informed feature engineering captures universal hardware bottlenecks rather than framework-specific artifacts.

\end{itemize}

\textbf{Accuracy \& Interpretability Trade-off.} We evaluated the system using real production data. As shown in Table~\ref{accuracy_comparison}, we compared the Full Feature strategy against our Physically-Informed Feature Engineering strategy using both Polynomial and GBDT models.

Although the Full Feature + GBDT combination yields the highest raw accuracy ($R^2=0.989$), it relies on software implementation details rather than hardware principles. As shown in the "Dominant Features" column, its top predictors are \texttt{post\_} features (e.g., \texttt{post\_MaxInLen}). In SGLang's architecture, these features correspond to the \texttt{process\_batch\_result} stage, which is overlapped with the current \texttt{run\_batch} computation. The model's reliance on these features indicates it is overfitting to the spurious correlations within the scheduling pipeline rather than the fundamental hardware bottlenecks.

In contrast, our Feature Engineering + GBDT strategy achieves competitive accuracy ($R^2=0.963$) while maintaining strict physical causality. It correctly identifies \texttt{Workload\_KV} as the primary determinant, aligning with the theoretical memory bandwidth bottleneck. This confirms that our model learns universal hardware constraints robust to software changes.

\textbf{Anomaly Detection Strategy.} Lacking external tools with comparable granularity, we evaluated three detection strategies by manually injecting faults: Fixed-Point (Fixed threshold 15\% + Point detection), Fixed-Window (Fixed 15\% + Window smoothing $W=10$), and Dynamic-Window (Dynamic $3\sigma$ + Window smoothing $W=10$).

As shown in Table~\ref{anomaly_strategy}, the Fixed-Window baseline achieved a perfect FPR (0.00\%), indicating that the injected faults were statistically distinct enough to trigger a conservative fixed threshold.
However, relying on a "magic number" (e.g., 15\%) is impractical for diverse production workloads. We adopt the Dynamic-Window strategy as the system default. Although it incurs a negligible FPR trade-off (0.59\%), it eliminates manual threshold tuning and maintains a higher recall (0.999 vs 0.993), ensuring robust detection across shifting noise baselines.

\begin{table}[!b]
\centering
\small
\caption{Performance Evaluation of Anomaly Detection Strategies on Production Datasets.}
\resizebox{\columnwidth}{!}{
\begin{tabular}{lccccc}
\toprule
\textbf{Strategy} & \textbf{Precision} & \textbf{Recall} & \textbf{F1} & \textbf{FPR} & \textbf{Lag} \\
\midrule
Dynamic-Point & 0.960 & 0.996 & 0.978 & 0.84\% & 0.0 \\
\textbf{Dynamic-Window} & \textbf{0.971} & \textbf{0.999} & \textbf{0.985} & \textbf{0.59\%} & \textbf{0.2} \\
Fixed-Window & 1.000 & 0.993 & 0.997 & 0.00\% & 1.4 \\
\bottomrule
\end{tabular}
}
\label{anomaly_strategy}
\end{table}

\begin{figure}[!b]
\centering
\includegraphics[width=\columnwidth]{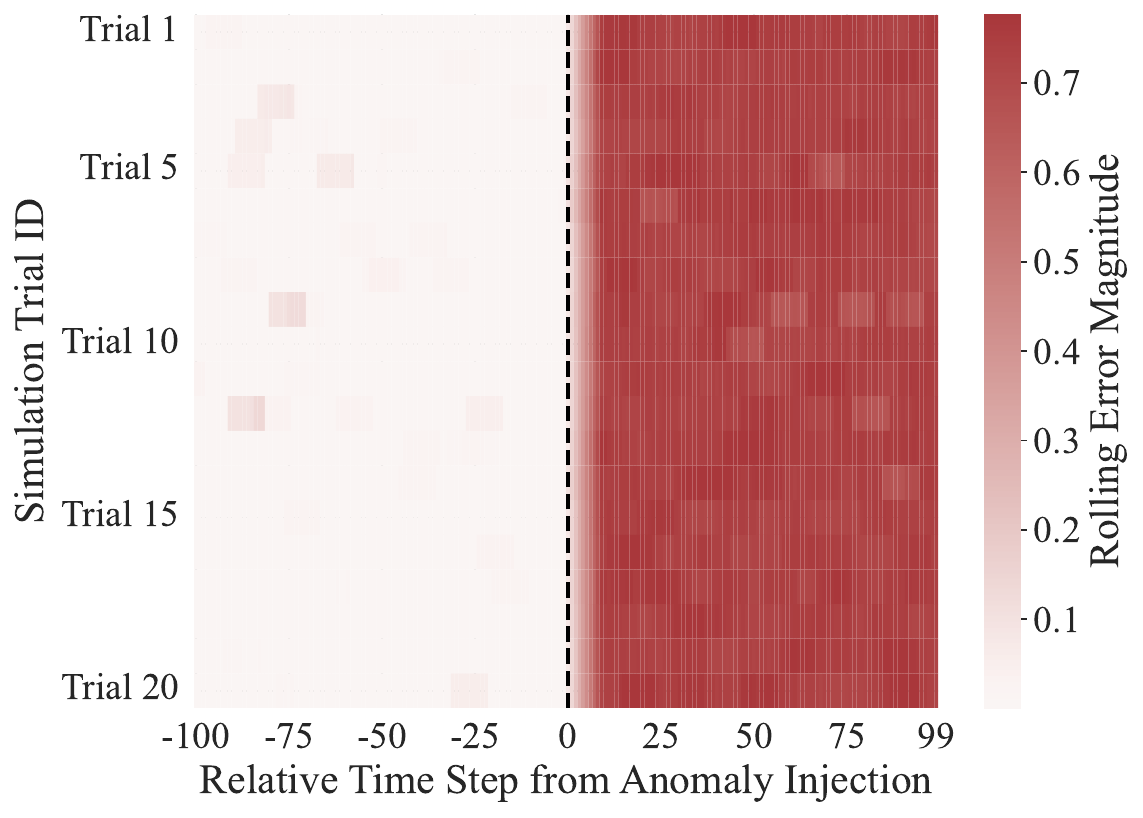}
\caption{Heatmap visualization of anomaly detection consistency across 20 independent trials using Dynamic-Window strategy. The dashed line marks the fault injection point. Red areas indicate detected.}
\label{fig:anomaly_heatmap}
\end{figure}

We visualize the detection performance across 20 independent simulation trials in Figure~\ref{fig:anomaly_heatmap}. As observed, there is a sharp, uniform transition from white (normal state) to deep red (high error) immediately following the anomaly injection point (dashed line). This visual evidence confirms that our strategy achieves low-latency detection and high consistency regardless of the specific random noise patterns in each trial, validating its suitability for real-time monitoring.

\subsection{Deep Dive Mode Assessment}

To assess the capability of \sysname{} in actual operations, we implemented a lightweight automated analysis script based on collected trace data. We emphasize that this root cause localization method is an exploratory downstream application instead of a core built-in component, demonstrating that the collected data is rich enough to support fine-grained fault reasoning.

\subsubsection{Metrics and Correlation}
Adapted from PerfTracker~\cite{2506.08528}, we utilize two primary metrics: \textbf{$\beta$ (Time Proportion)}, the percentage of cycle time consumed by an operation, and \textbf{$\mu$ (Average Utilization)}, the weighted average of associated hardware resources (e.g., GPU SM, CPU) during execution. We handle sparse counter samples via linear interpolation.

\textbf{Cycle-based Paradigm.} We adopt a cycle-based analysis rather than global averaging to capture transient dynamics inherent in LLM iterations. On the one hand, the diversity of captured events precludes a static priority order. On the other hand, we leverage the Python timeline to define strict cycle boundaries, where any function with high $\beta$ should be treated as a potential bottleneck. Therefore, unlike PerfTracker, we omit critical path analysis for $\beta$.

\subsubsection{Suspicion Score Algorithm}
Our algorithm assumes behavioral symmetry across distributed processes (e.g., within a Tensor Parallelism group). It identifies anomalies where $\beta$ exhibits statistically significant shifts. We utilize Z-Score ($Z = (\bar{x}_{abn} - \bar{x}_{norm}) / \sigma_{norm}$) to measure deviation. For resource utilization $\mu$, which follows a non-negative skewed distribution, we apply a log transformation ($\text{log1p}$) for robustness. The final suspicion score is defined as:
$
\text{Score} = |\Delta \beta| \times (|Z_{\beta}| + |Z_{\log \mu}|)
$.

\subsubsection{Validation}
We validated the method on SGLang and vLLM frameworks across CPU, GPU, and interconnect faults. As shown in Table~\ref{tab:root_cause}, the system successfully localizes root causes. All Top-1 anomaly cases exhibited P-values near 0.000, confirming that the collected traces provide statistically distinguishable signals for fault reasoning. Utilizing the resolved topology from §~\ref{sec:topology}, we can easily identify the straggler.

\begin{table}[!b]
\centering
\small
\caption{Root Cause Localization sample on SGLang.}
\label{tab:root_cause}
\resizebox{\columnwidth}{!}{
\begin{tabular}{llccrc}
\toprule
\textbf{Fault} & \textbf{Event} & \textbf{$\Delta\beta$ (\%)} & \textbf{Metric} & \textbf{$\Delta\mu$} & \textbf{Score} \\
\midrule
CPU Contention & \texttt{oncpu} & +29.0 & \texttt{cpu\_usage} & +50.1 & 2,821 \\
CPU Frequency Drop & \texttt{oncpu} & -6.4 & \texttt{frequency} & -63.2 & 177 \\
GPU Instability & \texttt{ampere\_gemm} & +5.9 & \texttt{gpu\_usage} & +15.3 & 1,561 \\
NVLink Congestion & \texttt{sglang::reduce} & +86.2 & \texttt{tx\_bytes} & +2.5e13 & 292k \\
\bottomrule
\end{tabular}
}
\end{table}

\section{Related Work}

\textbf{Hardware Vendor Tools.} Vendor-specific profilers such as NVIDIA Nsight Systems (\texttt{nsys})~\cite{nsight} and AMD \texttt{rocprof}~\cite{rocm_systems} offer authoritative, nanosecond-level visibility into GPU kernels and driver activities. While invaluable for offline kernel optimization, these tools are ill-suited for online production monitoring due to high overhead and requirement for invasive process launches. They also lack high-level application semantics. While extensions like NVTX allow manual instrumentation, this requires intrusive code modification.

\textbf{Framework-Level Tools.} These tools are deeply integrated into AI frameworks~\cite{TorchProfiler}. They understand internal logic and operator lifecycles, offering detailed high-level semantics. However, usage often requires wrapping code blocks with context managers, necessitating code modification and restarts. They also suffer from significant overhead~\cite{tensorflow} and are blind to events outside the framework, such as resource contention or scheduling latency.

\textbf{eBPF-based Tools.} Utilizing eBPF allows for zero code modification and no restarts with extremely low kernel-level overhead. eBPF naturally supports distributed tracing. While highly effective for CPU-bound microservices, standard eBPF probes cannot easily trace asynchronous GPU kernel execution, which is inadequate for deep diagnosis. For example, DeepFlow~\cite{deepflow} only provides limited CUDA insight for enterprise version~\cite{deepflowfeatures}. While eGPU\cite{eGPU} provides a powerful generic infrastructure for dynamic instrumentation, it remains agnostic to workload semantics, unable to transform raw monitoring signals into actionable SLO insights.

\textbf{Fine-grained Kernel Profiling.} Neutrino~\cite{neutrino} introduces a programmable probing framework operating at the assembly level (PTX/GCNAsm). It excels at capturing intra-kernel micro-behaviors. However, distributed LLM anomalies often stem from complex cross-stack interactions rather than isolated kernel inefficiencies. While Neutrino provides a powerful microscope for kernel debugging, \sysname{} provides the necessary cross-stack visibility to correlate high-level business semantics with low-level hardware signals across distributed nodes.

\textbf{Reliability for Distributed AI Workloads.} These systems target distributed training reliability. However, their localization precision is limited; for example, Mycroft~\cite{mycroft} cannot distinguish between unexpected performance issues and intentional overlap optimizations. Furthermore, Mycroft relies on instrumenting the NCCL library, imposing deployment and maintenance costs. Minder~\cite{minder} relies on metric thresholds, unable to distinguish root causes. Another example is XPUTimer~\cite{xputimer}, which also requires code modification.
\newline \newline
 We noticed a recent work on LLM inference lartency prediction called SCORPIO~\cite{scorpio}, which relies on linear latency assumptions ($Latency \propto Sequence Length$). Our experiments show that disabling SGLang overlap optimization do satisfy linear models, but it drops Prefill throughput by over 40\%, which is unacceptable. \sysname{} embraces production complexity by employing non-linear modeling (GBDT) to accommodate production optimizations, ensuring high-fidelity detection without forcing service degradation.

\section{Discussion}

\textbf{Interpretability vs. Black-Box Precision.} While high-dimensional raw features yielded slightly higher theoretical accuracy ($R^2 > 0.98$), we deliberately prioritized a sparse, physically-grounded feature set. This decision reflects a critical lesson in production operations: \textit{Actionability outweighs raw Accuracy}. A black-box alert indicating "Anomaly Detected (Score: 0.99)" breeds operator anxiety; a physically-informed alert pointing to "Decode Memory Bandwidth Saturation" empowers immediate remediation. By embedding hardware principles into the detection model, \sysname{} transforms monitoring signals into diagnostic logic, effectively extending the operator's cognition rather than replacing it.

\textbf{Adapting to Production Reality.} A fundamental tenet of \sysname{} is that observability must be a passive observer, not an active constraint. Modern inference engines rely on aggressive optimizations—Continuous Batching, Speculative Decoding, and computation-communication overlap—to maintain economic viability. Disabling these features to simplify modeling (e.g., forcing serialization) is unacceptable in production. \sysname{} accepts the non-deterministic reality of these optimizations. Instead of demanding a sanitized environment, we engineered robust mechanisms to extract signal from the noise of high-performance execution.

\textbf{Mechanism-Policy Separation in Diagnosis.} \sysname{}'s core deliverable is structured, cross-stack aligned execution context. By providing high-fidelity, semantically aligned trace data as a neutral substrate, \sysname{} enables efficient human intervention by ensuring time spent on solving verified problems rather than sifting through noise. The mechanism we presented is an independent downstream tool to validate data richness. We have not integrated it into the main system because hard-coding rigid diagnostic rules into the monitoring agent invites brittleness under highly dynamic environments. This decoupling allows downstream users to develop specialized diagnostic applications tailored to evolving production needs without bloating the core infrastructure.

\textbf{Limitations and Future Work.} While \sysname{} excels in latency spike detection, limitations exist. Its monitoring model relies on historical normal data for baseline construction. Therefore, new workloads or hardware upgrades require a brief warm up. Pre-existing persistent performance issues may not trigger alerts due to the lack of a valid normal reference. Semantic correlation for non-mainstream XPUs is limited by vendor toolchain openness. Future work will explore leveraging LLMs for natural language summarization of traces and developing calibration-free adaptive baseline mechanisms.

\section{Conclusion}

As Large Language Models transition from research artifacts to critical utility services, the operational focus must shift from training throughput to inference stability. We present \sysname{}, the first zero-intrusion, full-stack performance tracking and anomaly detection system for production LLM inference. \sysname{} solves the long-standing disconnection between upper-layer business logic and underlying physical resource states. Through large-scale deployment and controlled experiments, we validated its effectiveness: it maintains negligible overhead ($<$0.5\% CPU), achieves precise anomaly detection (F1=0.985) via dynamic baseline modeling, and provides rich, alignable trace data that accelerates root cause analysis. \sysname{} demonstrates that building an intelligent, transparent, and efficient observability system is essential for guaranteeing service quality and unlocking computational potential in complex AI infrastructures. Furthermore, we release the first open-source anomaly simulation suite, lowering the barrier for future research to investigate transient performance regressions in complex AI infrastructures.

\section*{Acknowledgments}

This work was supported by Alibaba Research Intern Program.

\section*{Availability}
We provide the source code of core logic, anomaly injection scripts, and the experimental results at \url{https://github.com/kunluninsight/LatencyPrism}.
\bibliographystyle{plain}
\bibliography{main}

\end{document}